\documentclass[12pt,tightenlines,eqsecnum,floats,showpacs,nofootinbib,amsmath,amssymb,aps,prd]{revtex4}

\usepackage{graphicx}
\usepackage{amsmath,amssymb}
\usepackage{setspace}
\usepackage{amsmath,amssymb,amsfonts,amsthm}
\usepackage{graphicx}
\usepackage{enumerate} 
\usepackage{colordvi} 
\usepackage{subfigure}

\def\be{\begin{equation}}
\def\ee{\end{equation}}
\def\ba{\begin{eqnarray}}
\def\ea{\end{eqnarray}}

\def\f{\frac}
\def\B{\mathcal{B}}
\def\L{\mathcal{L}}
\def\H{\mathcal{H}}
\def\Hgrav{\H_{\rm grav}}
\def\Hmatt{\H_{\rm matt}}
\def\Heff{\H_{\rm eff}}
\def\Hkin{\H_{\rm kin}}
\def\a{\alpha}
\def\rcr{\rho_{\rm crit}}
\def\pl{{\rm Pl}}
\def\lp{\ell_{\rm Pl}}
\def\SU(2){\rm SU(2)}
\def\e{\underbar{e}}
\def\w{\underline{\omega}}

\begin{document}


\title{The covariant entropy bound and loop quantum cosmology}

\author{Abhay Ashtekar}
\email{ashtekar@gravity.psu.edu}
\author{Edward Wilson-Ewing}
\email{wilsonewing@gravity.psu.edu} \affiliation{Institute for
Gravitation and the Cosmos \& Physics Department, Penn State,
University Park, PA 16802, U.S.A.}

\begin{abstract}

We examine Bousso's covariant entropy bound conjecture in the
context of radiation filled, spatially flat,
Friedmann-Robertson-Walker models. The bound is violated near the
big bang. However, the hope has been that quantum gravity effects
would intervene and protect it. Loop quantum cosmology provides a
near ideal setting for investigating this issue. For, on the one
hand, quantum geometry effects resolve the singularity and, on the
other hand, the wave function is sharply peaked at a quantum corrected
but smooth geometry which can supply the structure needed to test
the bound. We find that the bound is respected. We suggest that the
bound need not be an essential ingredient for a quantum gravity
theory but may emerge from it under suitable circumstances.

\end{abstract}


\maketitle

\section{Introduction}
\label{s1}

Over 30 years ago, Bekenstein made the seminal suggestion that
black holes carry an entropy equal to a quarter of their surface
area in Planck units \cite{bek1, bek2}. The discovery of the first
law of black hole thermodynamics by Bardeen, Carter and Hawking
\cite{bch,akrev}, coupled with Hawking's discovery of black hole
radiance \cite{swh} made this suggestion compelling. The ensuing
generalized second law of thermodynamics \cite{bek3} in turn
motivated the more recent holographic principle by 't~Hooft and
Susskind \cite{'tH, suss} as well as several entropy bound
conjectures. It has been suggested that the holographic principle is
a powerful hint and should be used as an essential building block
for any quantum gravity theory, much as the principle of equivalence
is fundamental to general relativity \cite{'tH, suss, bousso2}.

Perhaps the most promising of the entropy bound conjectures is due to
Bousso \cite{bousso1, bousso2}. It is formulated in precise
geometric terms and limits the entropy flux through a surface's
``light-sheet'' by the area of that surface. More precisely, let $A$
be the area of an arbitrary spatial 2-surface $\B$. A
three-dimensional null hypersurface $\L$ is a light-sheet of $\B$ if
it is generated by nonexpanding light rays that begin at $\B$ and
extend orthogonally away from $\B$.  Let $S$ denote the total
entropy flux of the matter fields across any light-sheet $\L$ of $\B$.
Then (in units where c=k=1)
\be \label{conj} S \,\,\le\,\, \f{A}{4G\hbar}. \ee
In general, $S$ is difficult to compute unless an entropy current $s^a$
can be introduced, in which case
\be S=\int_{\L}\, s^a \epsilon_{abcd}, \ee
where $\epsilon_{abcd}$ is the space-time volume 4-form. Although
the existence of an entropy current is not necessary for the bound
to be meaningful, it often is not clear how to compute $S$ if $s^a$
does not exist.

The conjecture has several curious features. First, for it to be
easily tested, matter should admit an entropy current. This is a
strong requirement since a well-defined expression of entropy
current is generally not available unless the matter can be
represented as a fluid. Therefore, standard space-times with
``fundamental'' matter fields ---such as scalar, Maxwell, and gauge
fields--- that one often considers in general relativity cannot
readily be tested. Thus, circumstances where the conjecture can be
tested are rather limited. However, it is still a highly
nontrivial question as to whether the bound holds for matter
sources that do admit a well-defined entropy current. Remarkably,
the answer is in the affirmative provided the entropy current and
stress-energy tensor satisfy certain inequalities along $\L$
\cite{fmw, bfm}. Moreover, these inequalities can be motivated
from statistical physics of standard bosons and fermions provided
one remains away from the Planck scale. Finally, there is also a
necessary and sufficient condition \cite{ap} for the validity of
the bound, based on a plausible assumption on the time scale on
which thermodynamic equilibrium can be reached.

A second feature of the conjecture is that its current proofs
\cite{fmw,bfm} require that matter fields satisfy the dominant (or
at least null) energy condition; otherwise it would be possible to
add entropy across $\L$ without adding energy and appreciably
changing the area of $\B$. This assumption is completely
reasonable within classical general relativity and indeed
constitutes a building block of many of the hard results in this
theory. However, since the right side of (\ref{conj}) becomes
infinite in the limit $\hbar$ goes to zero, the conjecture is
nontrivial only in the realm of quantum physics. But in quantum
field theory, expectation values of the matter stress-energy
tensor fail to satisfy the dominant (or the null) energy
condition. Thus there is a clear tension between the classical
formulation of the covariant entropy bound and the quantum world
it is trying to represent.  What would happen in full quantum
gravity? Would a suitable generalization of the conjecture
survive? As remarked above, it has been suggested that an
appropriate generalization should in fact be a building block of
any quantum gravity theory.

Finally, as we will see in Sec. \ref{s2}, even in the k=0,
radiation dominated Friedmann-Robertson-Walker (FRW) space-time
---a classical solution of Einstein's equation, which admits a
perfect fluid as source--- the conjecture is violated near the big
bang.%
\footnote{Such problems in the deep Planck regime very near the
big bang were encountered also in the earlier versions of entropy
bounds \cite{fs}.}
(In this regime the inequalities assumed in \cite{fmw,bfm,ap}
fail.) It is natural to suppose that this violation is spurious,
because quantum gravity effects would be dominant in this regime
and modify the dynamics of general relativity. But one's first
expectation would be that in this Planck regime the space-time
metric would fluctuate violently,%
\footnote{Explicit examples of such phenomena are provided by some
symmetry reduced models. See, e.g., \cite{aa-large, atv}.}
making it impossible to speak of a spacelike surface $\B$ and
especially its light-sheet $\L$. If so, the basic ingredients in the
very statement of the conjecture would be unavailable. However, loop
quantum cosmology (LQC) provides a rather unexpected situation
\cite{aps-lett,aps2,apsv}: Einstein's equations do get modified, the
classical singularity \emph{is resolved} and replaced by a quantum
bounce, but all this physics is well captured by a quantum corrected
metric, which is again a smooth tensor field (although its expression
contains $\hbar$ corrections, which dominate the dynamics near the
bounce). Therefore, the structure required for Bousso's formulation
is available. It is then natural to ask: Does the bound hold?
We will show that the answer is in the affirmative.

This result by itself is not a strong statement in favor of either
the covariant bound or LQC. Had it been violated, one could have
argued that the fluid approximation is inappropriate near the bounce
where the matter density is close to the Planck density, or that
LQC, with its focus on homogeneous and isotropic situations, is too
restrictive. Nonetheless, the fact that there is a coherence between
ideas that were independently developed from completely different
motivations and perspectives is quite striking. There may well be a
deeper underlying reason behind this apparent unity. Our viewpoint,
motivated by results of loop quantum gravity (LQG), is the
following. In the Planck regime, space-time geometry is quantum
mechanical and has a fundamental, in-built discreteness (see, e.g.,
\cite{alrev,crbook,ttbook}). This discreteness is directly
responsible for many of the novel features of the theory, including
the resolution of the big bang singularity through a quantum bounce
\cite{aps-lett,aps2,apsv}. Because of this discreteness, the true
degrees of freedom of nonperturbative quantum gravity are very
different from those of standard quantum field theories on
background space-times. This fact has important consequences
which have a ``holographic character'' ---e.g. the derivation of the
entropy of black holes and cosmological horizons
\cite{abck-lett,abk,aev}. Bousso's covariant entropy bound could thus
emerge from LQG in suitable circumstances, including some in which
the assumptions made in \cite{fmw,bfm,ap} are violated. Thus, rather
than being an essential building block for quantum gravity, the
bound could \emph{emerge} from a background independent,
nonperturbative theory which appropriately incorporates the quantum
nature of geometry in the Planck regime.

The paper is organized as follows. In Sec. \ref{s2}, we show that
in the k=0, radiation filled FRW cosmology,  the covariant entropy
bound is violated near the big bang singularity. In Sec. \ref{s3},
we first recall the basic results from LQC and then show that the
bound is in fact respected in the quantum corrected ``effective''
space-time geometry that descends from LQC. In the classical as well
the LQC analysis of the bound, we only consider ``round'' 2-spheres
$\B$. As in other discussions ---e.g., of marginally trapped
surfaces--- in general relativity, treatment of general surfaces
would increase the technical complexity very significantly. Section
\ref{s4} summarizes the results and discusses related issues.

\section{FRW Universe}
\label{s2}

The line element of a flat FRW universe is given by
\be\label{metric} ds^2 = -d\tau^2 + a(\tau)^2 \left(dr^2 + r^2
d\Omega^2\right), \ee
and the Friedmann equations describing the evolution of this
universe are
\be \label{fried} \left(\f{\dot{a}}{a}\right)^2 = \f{8\pi G}{3}\rho,
\ee
\be \label{cont} \dot{\rho} + 3\f{\dot{a}}{a}\left(\rho+P\right)
=0.\ee
We will restrict ourselves to a radiation-dominated universe where the
required entropy current can be constructed using standard
statistical mechanical considerations. The equation of state
$P=\f{1}{3}\rho$ of the fluid gives the solution
\be \label{aFRW} a(\tau) = \left(\f{32\pi
GK_o}{3}\tau^2\right)^{1/4} \quad\quad \mathrm{and}\quad\quad
\rho(\tau) = \f{K_o}{a(\tau)^4}, \ee
where $K_0$ is an integration constant without direct physical
significance, which drops out of the expressions of measurable
quantities.

We will take the radiation fluid to be a photon gas and assume that
the universe is always in instantaneous equilibrium. Then, from
standard statistical mechanics we have
\be \label{therm1} \rho=\f{\pi^2}{15\hbar^3}\,T^4, \ee
where $T$ is the temperature. From (\ref{aFRW}) we then conclude as
usual that the temperature has the following time dependence:
\be T=\left(\f{45\hbar^3}{32\pi^3G}\right)^{1/4}\f{1}{\sqrt{\tau}}.
\ee

The entropy density $s$ of a photon gas is given by
\be \label{therm2} s=\f{\rho+P}{T}. \ee
Therefore,  the entropy flux through a light-sheet $\L$ is given by
\be S = \int_{\L} s^a\epsilon_{abcd}, \ee
where the entropy density 4-vector is $s^a= su^a$ with $u^a$, the
unit vector normal to constant $\tau$ slices and where, as before,
$\epsilon_{abcd}$ is the space-time volume 4-form. For our
calculations, we will choose our spatial 2-surface $\B$ to be a
metric 2-sphere at time $\tau_f$. Since the FRW space-time does not
admit a trapped surface, $\B$ must admit a past light-sheet.
Furthermore, since the FRW space-times are conformally flat and $\B$
is a round 2-sphere, $\L$ would either terminate on the big bang
singularity or be the future null cone of a point. Without loss of
generality we can assume that $\L$ is generated by the null vector
$k^a=(-1,-\f{1}{a},0,0)$. Then, if $\L$ is the null cone of a point
at $\tau=\tau_i > 0$, we find that
\be \label{S} S = \f{16\pi}{9}\left(\f{\pi^2
K_o^3}{15\hbar^3}\right)^{1/4}R_f^3, \ee
where
\be \label{Rf} R_f = \int_{\tau_i}^{\tau_f}
\f{d\tau^\prime}{a(\tau^\prime)} \ee
is the radius of $\B$. $R_f$, and hence $S$, is well defined so
long as $\tau_i>0$. Since the area of $\B$ is simply
\be \label{A} A=4\pi \: a(\tau_f)^2R_f^2\, , \ee
we obtain
\be \f{S}{A}=\f{1}{6}\left(\f{2}{45\pi
G^3\hbar^3}\right)^{1/4}\f{1}{\sqrt{\tau_f}}
\left(1-\sqrt{\f{\tau_i}{\tau_f}}\right) \ee
as the ratio of the entropy flux through $\L$ to the area of $\B$.
Clearly this ratio diverges as $\tau_f$ approaches zero (keeping
$\tau_i < \tau_f$). Therefore, one can violate the covariant entropy
bound by an arbitrary amount by choosing $\B$ sufficiently close to
the big bang.

However, by plugging in numbers it is easy to show that the bound
does hold if $\tau_f \gtrsim 0.06 t_{\pl}$ or, equivalently, the
matter density satisfies $\rho \lesssim 8.3\rho_{\pl}$. Thus, the
bound is violated only in the deep Planck regime where quantum
gravity effects are expected to be dominant. Indeed, since the
phenomenological fluid approximation is likely to fail for much
larger values of $\tau_f$, it is surprising that the bound has
such a large domain of validity! Nonetheless, since it does fail,
it is natural to ask whether the quantum gravity effects that
resolve the big bang singularity can also restore the validity of
the bound. In the next section, we will analyze this issue in the
context of LQC and show that the answer is in the affirmative.

\section{Quantum bounce and the entropy bound}
\label{s3}

This section is divided into two parts. In order to make the paper
self-contained, in the first part we provide a concise summary of
ideas that underlie LQC and results that have emerged from it,
focusing on those features that are most relevant to our main
result. Readers who are already familiar with LQC will find the
streamlined summary of ``effective'' equations useful. Readers who are
interested only in the entropy bound can skip this material and
go directly to the next subsection where we analyze the covariant
entropy bound in the quantum corrected, ``effective'' space-time that
emerges from LQC.

\subsection{Loop Quantum Cosmology}
\label{s3.1}

Loop quantum cosmology is an application of the techniques of LQG
\cite{alrev,crbook,ttbook} to space-times
that are homogenous and isotropic. Currently, symmetry reduction is
carried out at the classical level; LQC is yet to be derived
systematically from LQG. However, one quantizes the reduced system
by mimicking full LQG as closely as possible. Therefore, the
mathematical structure of LQC closely resembles that of full LQG,
thus differing from the older Wheeler-DeWitt quantum cosmology
\emph{already at the kinematical level}.%
\footnote{For a relatively short review of LQC and its relation to
LQG,  see, e.g., \cite{aa-rev}.}

In LQG the gravitational phase space is the same as that in the
$\SU(2)$ Yang-Mills theory. Thus, the configuration variable is an
$\SU(2)$ connection $A_a^i$ and its conjugate momentum is a vector
density $E^a_i$ of weight 1 which also takes values in the Lie
algebra of $\SU(2)$.%
\footnote{Thus, indices $a,b,c,\ldots$ refer to the tangent space of
the ``spatial'' 3-manifold and $i,j,k,\ldots$ to the Lie algebra of
$\SU(2)$.}
However, $A_a^i$ is now the \emph{gravitational} connection used to
parallel propagate chiral spinors. Similarly, the ``electric field''
$E^a_i$ now has a direct geometrical interpretation: it represents a
(density weighted) orthonormal triad that determines the spatial
Riemannian geometry. The fundamental quantum algebra $\mathfrak{a}$
is generated by holonomies $h_e$ defined by the gravitational
spin-connection $A_a^i$ along (1-dimensional curves or) edges $e$,
and fluxes $E_S$ of the ``electric fields'' $E^a_i$ across 2-surfaces
$S$. The key task in quantum kinematics is to find an appropriate
representation of $\mathfrak{a}$ on a Hilbert space $\Hkin$.
Somewhat surprisingly, it turns out that the requirement of
background independence selects the representation uniquely
\cite{lost}! This representation provides the arena to formulate
quantum dynamics. Now, the classical dynamics of the gravitational
and matter fields is generated by a set of first class constraints.
These classical constraints have to be promoted to well-defined
self-adjoint operators on $\Hkin$. Finally, the physical Hilbert
space is built from suitable solutions to these operator
constraints. As in any background independent system, these physical
states already encode quantum dynamics which can be made explicit by
choosing one of the degrees of freedom
---such as a matter field or, in cosmology, the scale factor--- as
an internal clock with respect to which other degrees of freedom
evolve.

When one symmetry reduces general relativity by imposing homogeneity
and isotropy, it turns out simplest to gauge fix and solve the
so-called Gauss and vector constraints, which respectively generate the
internal $\SU(2)$ rotations of triads and spatial diffeomorphisms.
One is then left just with the Hamiltonian constraint, which
generates evolution (in proper time). In the k=0, FRW model, it is
given by
\ba \Hgrav + \Hmatt = 0,\quad &{\rm where}& \nonumber\\
\Hgrav = - \,\gamma^{-2}\, \epsilon^{ij}_{\phantom{ij}k}
\f{E^a_iE^b_j}{\sqrt{E}}\, F_{ab}^k,\quad  &{\rm and}& \quad \Hmatt=
\rho \sqrt{E}\, , \ea
where $\gamma$ is the so-called Barbero-Immirzi parameter,
$F_{ab}^k$ the field strength of the connection $A_a^i$,  $\rho$,
the matter density and $E$, the determinant of $E^a_i$ (or
equivalently, of the spatial metric $q_{ab}$ determined by $E^a_i$).
Let us focus on the gravitational part $\Hgrav$ of this constraint.
The term containing triad can be taken over to a quantum operator in
a natural fashion, first spelled out by Thiemann \cite{tt,ttbook}.
Therefore, the key problem in passage to the quantum theory lies in
the construction of an operator $\hat{F}_{ab}^k$ representing field
strength. For, while the unique representation of $\mathfrak{a}$
selected by the requirement of background independence admits an
operator $\hat{h}_e$ representing holonomies, these fail to be
continuous in the edge $e$, whence there is no operator
corresponding to the connection $A_a^i$ itself. The strategy then is
to use holonomy operators to obtain $\hat{F}_{ab}^k$.

Now, it is well known that in classical geometry, the field strength
can be recovered as the limit of the ratio of the holonomy around an
appropriate closed loop divided by the area of the loop, as one
shrinks the loop thereby sending its area to zero. We can use the
same idea for quantization of $F_{ab}^k$. However, now because of
quantum geometry, eigenvalues of the area operator are discrete.
Therefore, to define the action of $\hat{F}_{ab}^k$ on a given LQC
state, we are led to shrink the loop only until the physical area it
encloses reaches the minimum nonzero eigenvalue, say $\a_o\lp^2$
(in the class of LQG states compatible with the LQC state under
consideration). The resulting $\hat{F}_{ab}^k$ is a self-adjoint
operator on the kinematical Hilbert space $\Hkin^{\rm LQC}$ with a
built-in nonlocality at a scale $\a_o$. The viewpoint in LQC is
that this nonlocality is fundamental and the more familiar, local
expression of the field strength arises only in the classical limit.
(For details, see \cite{abl,aps2,aa-rev}.) This nonlocality is
important for quantum dynamics: it is the principal reason why the
LQC Hamiltonian constraint is qualitatively different from the
Wheeler-DeWitt differential equation.

Solutions to the LQC Hamiltonian constraints were first obtained
numerically \cite{aps2} and an analytical understanding was
developed more recently \cite{acs}. If one begins with a suitable
quantum state that is sharply peaked on a classical trajectory at
late times and evolves it both forward and backward in time, the
trajectory remains sharply peaked for all times; the theory admits
``dynamical coherent states''. However the trajectory on which they
remain peaked are classical solutions only till the matter density
$\rho$ reaches about 1\% of the Planck density $\rho_{\pl}$. Then,
the quantum evolution departs from the classical trajectory. Rather
than evolving into the singularity in the backwards evolution, the
trajectory undergoes a quantum bounce. The density then starts
decreasing again. Once it falls below 1\% of $\rho_{\pl}$, classical
general relativity again becomes an excellent approximation. Thus,
the effect of quantum geometry is to produce an effective
\emph{repulsive force} which is negligible till $\rho \sim 0.01
\rho_{\rm Pl}$ but rises \emph{extremely quickly} once the density
further increases, so much so that it is able to overwhelm the
classical gravitational attraction, thereby triggering the bounce
and avoiding the classical singularity. It is this short but
critical interval that will be important for our entropy
considerations in the next subsection.

The LQC quantum evolution is extremely well approximated by quantum
corrected ``effective'' equations. These are obtained \cite{jw} using
a geometrical formulation of quantum mechanics.%
\footnote{In this formulation, the space of quantum states is
represented by an infinite dimensional phase space $\Gamma_{\rm
Quan}$ on which the exact quantum dynamics defines a Hamiltonian
flow (see, e.g., \cite{as}). The quantum corrected equations are
obtained by projecting this flow on a suitably chosen subspace
which is isomorphic to the classical phase space $\Gamma_{\rm Cl}$.
To distinguish them from effective equations obtained from other
methods, these equations should really be referred to as ``geometric
quantum mechanics projected equations''. However, for brevity we will
refer to them just as ``effective'' equations.}
We will conclude this subsection by sketching the structure of
these equations. Recall first that in the FRW models one generally
introduces a fiducial frame $\e^a_i$ and the corresponding
co-frame $\w_a^i$ and expresses the dynamical variables in terms
of them. Homogeneity and isotropy imply that the connection
$A_a^i$ is proportional to $\w_a^i$  ---\,\,$A_a^i\, \sim\,\, c\,
\omega_a^i$\,\,---  and the triad $E^a_i$ to $\e^a_i$
---\,\,$E^a_i\, \sim\,\, p\, \e^a_i$\,\,--- where, for simplicity
we have omitted some kinematical factors that are irrelevant for
dynamics. $c,p$ is then the basic canonically conjugate pair. The
kinematical factors make $c$ dimensionless, endow $p$ with
dimensions of ${\rm (length)}^2$, and provide the following
Poisson bracket: $\{c,p\} = 8\pi \gamma G/3$. The geometrical
meaning of these variables is the following: $c \sim \gamma
\dot{a}$ and $|p| \sim a^2$ where, as in (\ref{metric}), $a$ is
the scale factor (where, again, we have omitted some kinematical
proportionality factors that are irrelevant for dynamics). The
Hubble parameter is given by $H :={\dot{a}}/{a} = {c}/{\gamma
\sqrt{|p|}}$.

In terms of these variables the effective Hamiltonian constraint
turns out to be:
\be \label{heff} \Heff = - \f{3|p|^{3/2}}{8\pi G\gamma^2
\a_o\lp^2}\,\, \sin^2\, \sqrt{\f{\a_o\lp^2 c^2}{|p|}}  + |p|^{3/2}
\rho = 0\, .\ee
As in the classical theory, the canonical transformation generated
by this constraint yields evolution equations (in proper time).
However, $\Heff$ has a dependence on Planck's constant through
$\lp=\sqrt{G\hbar}$. The classical Hamiltonian constraint results
when we take the limit $\hbar \rightarrow 0$. Both $\a_o$ and the
Barbero-Immirzi parameter $\gamma$ disappear in this limit, as they
must. Note that only the gravitational part of $\Heff$ has acquired
quantum corrections; as one would expect, they arise from the
fundamental nonlocality of $\hat{F}_{ab}^k$ in LQC. However, to
compare with the standard form of the classical constraint ---the
Friedmann equation--- it is convenient to rewrite this equation in
a slightly different form. Using the equation of motion
\be \dot{p} := \{p, \Heff\}\, = \, \f{2|p|}{\gamma \a_o\lp^2}\, \sin
\sqrt{\frac{\a_o c^2\lp^2}{|p|}}\,\,\cos \sqrt{\frac{\a_o
c^2\lp^2}{|p|}} \ee
for $p$, the fact that $H^2 = {(\dot{p})^2}/{4p^2}$, and
(\ref{heff}) we obtain
\be \label{mfried} H^2 = \f{8\pi G\rho}{3}\, \left(1 -
\f{\rho}{\rcr}\right), \quad {\rm with} \quad \rcr =
\f{3}{8\pi\gamma^2 \a_o G \lp^2}. \ee
Thus, using the equation of motion $\Heff$ generates, we have moved
the quantum correction from the ``geometric'' to the ``matter'' side.
(This procedure is quite general and holds so long as $\Hmatt$ does
not depend on the gravitational connection $c$.) As $\hbar$ goes to
zero, $\rcr$ diverges and the second term in the parenthesis on the
right-hand side disappears; we recover the classical Friedmann
equation, Eq. (\ref{fried}). The quantum correction is negligible when
$\rho \ll \rcr \sim \rho_{\rm Pl}$. But it dominates dynamics when
$\rho \sim \rcr$. In particular, when $\rho =\rcr$ the right side
vanishes, whence $\dot{a} =0$ and we have a quantum bounce.

The expression of the critical density contains two dimensionless
parameters, $\gamma$ and $\a_o$. In LQG, the value of the
Barbero-Immirzi parameter $\gamma$ is fixed through a black hole
entropy calculation \cite{abck-lett,abk} and turns out to be
$\gamma \approx 0.2375$. The second parameter is $\alpha_o$. In
the LQC literature to date it has been taken to be the minimum
nonzero eigenvalue of the area operator, $\alpha_o =
2\sqrt{3}\pi\gamma$, on gauge invariant states (see, e.g.,
\cite{aps2,apsv,acs}). The corresponding area eigenstate has two
edges: an edge $e_1$, which terminates \emph{transversally} at a
vertex $v$ on the surface whose area is being computed, and meets
there an edge $e_2$ that is \emph{tangential} to the surface (each
edge carrying a spin-label $j=1/2$). However, we recently realized
that if one sets up a semiheuristic correspondence between LQG
and LQC quantum states, such states would not occur in homogeneous
models: at $v$, $e_1$ would meet an edge $e_2$, \emph{which is also
transversal} to the surface. Therefore in LQC one should take the
minimum area eigenvalue \emph{within this class}. That minimum is
$\alpha_o =
4\sqrt{3}\pi\gamma \approx 5.166$.%
\footnote{These considerations become quite important in a
systematic treatment of the Hamiltonian constraint in non-isotropic
models.}
Thus, in this paper we will set $\gamma= 0.2375$ and $\alpha_o =
5.166$.

\subsection{Entropy bound in the LQC-corrected FRW model}
\label{s3.2}

Since we have restricted ourselves to the k=0, isotropic homogeneous
situation, the effective space-time metric in LQC again has the form
\be \label{lqc-metric}
ds^2=-d\tau^2+a(\tau)^2\left(dr^2+r^2d\Omega^2\right). \ee
Any symmetric, second rank tensor field  $T_{ab}$ that is invariant
under these symmetries can be written as $T_{ab} = \rho \nabla_a\tau
\nabla_b \tau + P (g_{ab} \,+\, \nabla_a \tau \nabla_b\tau)$. Hence
the stress energy tensor is necessarily that of a perfect fluid.
Since we are working with a radiation filled universe, $P=\f{1}{3}\rho$.
As usual we can model it using kinetic theory on a curved
space-time. The resulting conservation equation $\nabla^aT_{ab}=0$
again yields the continuity equation:
\be \label{cont2} \dot{\rho}
+3\f{\dot{a}}{a}\left(\rho+P\right)=0.\ee
Thus the form of the metric and the continuity equations are the
same as in the classical theory. However as we saw in Sec.
\ref{s3.1}, the Friedmann equation is now replaced by:
\be \label{mfried2} H^2 = \f{8\pi G\rho}{3}\, \left(1 -
\f{\rho}{\rcr}\right), \quad {\rm with} \quad \rcr =
\f{3}{8\pi\gamma^2 \a_o G \lp^2}. \ee
We can repeat the procedure followed in Sec. \ref{s2} and use
these three equations to solve for $a(\tau)$. The result
is:
\be \label{aLQC} a(\tau)=\left(\f{32\pi
GK_o}{3}\tau^2+\f{K_o}{\rcr}\right)^{1/4}, \quad\quad
\mathrm{and}\quad\quad \rho(\tau) = \f{K_o}{a(\tau)^4} \ee
where, as before, $K_o$ is an integration constant which does not
have a direct physical significance. This functional form of
$a(\tau)$ is very similar to Eq. (\ref{aFRW}) in the classical case,
and reduces to it in the limit $\hbar \rightarrow 0$ as well as in
the limit $\tau \rightarrow \pm \infty$. The critical density $\rcr$
at which the universe bounces is a measure of the strength of the
effects of quantum gravity. The smaller $\rcr$ is (or, the larger
the area gap $\a_o$ is), the earlier the onset of the quantum
regime will be. Now, in the classical theory, the ratio $S/A$ could
be made arbitrarily large by choosing the surface $\B$ sufficiently
close to the singularity. However this is precisely the regime in
which LQC effects become dominant. Therefore we are now led to ask:
\begin{itemize}
\item Do these effects naturally bound the ratio $S/A$ under
    consideration?
\item If so, for the values of parameters $\gamma$ and $\alpha_o$ that are
currently used in LQG and LQC, is the bound less than $0.25/\lp^2 $?
\end{itemize}

These questions can now be answered by a straightforward
calculation. Let us again choose a round 2-sphere $\B$ at an instant
of time $\tau_f$ and consider its past lightsheet. Since the
space-time metric (\ref{lqc-metric}) is conformally flat \emph{and
nonsingular}, these null rays necessarily converge to a point $p$,
say at time $\tau_i$,  so that $\B$ is a cross section of the
(future) light cone of $p$.%
\footnote{It is not necessary to consider, in addition, past light
cones because the quantum corrected ``effective'' solution is
symmetric about the bounce point at $\tau=0$.}
However as one moves along the light cone from $\B$ to $p$, the
expansion may become positive somewhere, in which case the light
sheet of $\B$ would only be a \emph{portion} of the (future) light
cone of $p$. If not, the light-sheet is the entire light cone from
$p$ to $\B$. Both these possibilities occur. But in either case,
the area of $\B$ is just
\be A = 4\pi [a(\tau_f)]^2R_f^2\quad\quad {\rm where} \quad\quad
 R_f = \int_{\tau_i}^{\tau_f} \f{d\tau^\prime}{a(\tau^\prime)}\, .
 \ee
Using the space-time metric it is easy to calculate the
entropy current $s^a = [(\rho+P)/T]u^a$ through \emph{the full light
cone}, where $u^a$ is the unit normal to the $\tau={\rm const}$
slices. We obtain:
\be \f{S}{A} =
\f{4}{9}\,\,\left(\f{\pi^2K_o}{15\hbar^3}\right)^{1/4}\,\,\,
\f{R_f}{\sqrt{\f{32\pi G\tau^2_f}{3} + \f{1}{\rcr}}}, \ee
where $R_f$ is determined by a hypergeometric function:
\be R_f(\tau_i,\,\tau_f) =\left[\tau
\left(\f{\rcr}{K_o}\right)^{1/4}\, {}_2 F_1\left(\f{1}{2},
\f{1}{4};\f{3}{2}, -\f{32\pi
G\rcr}{3}\tau^2\right)\,\right]_{\tau_i}^{\tau_f}\, . \ee
If the light-sheet of $\B$ is only a portion of the light cone, the
$S/A$ relevant for the covariant entropy bound will be smaller.

\begin{figure}[tb]
  \begin{center}
    \includegraphics[width=4in,angle=0]{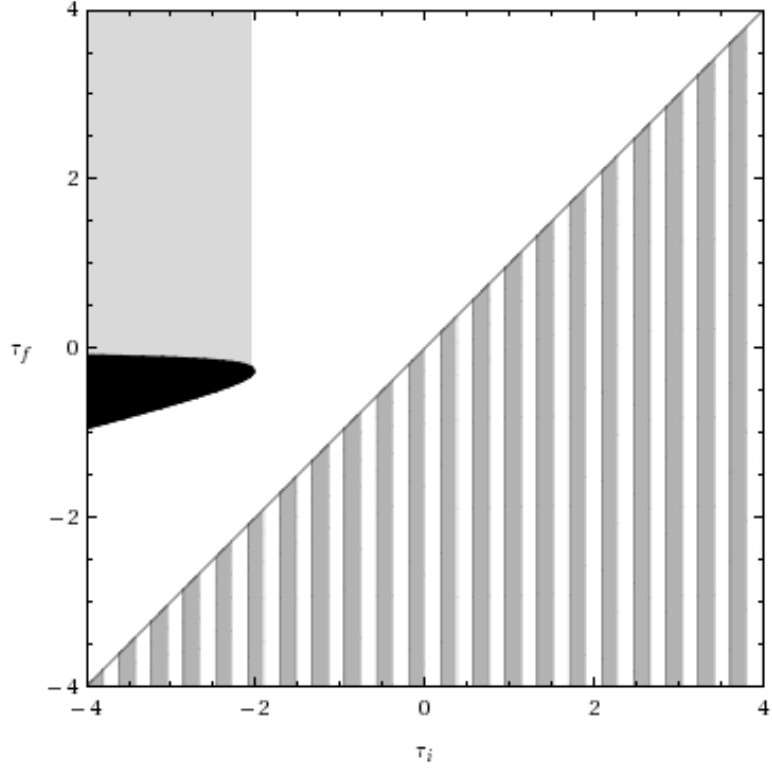}
    \caption{ The $(\tau_i, \tau_f)$ region (in Planck units) where LQC
    corrections are significant. Since $\tau_i <\tau_f$, the striped portion
    is excluded. For $(\tau_i,\tau_f)$ in the black region, surfaces $\B$ do
    not admit past light-sheets. In the grey region, the expansion is negative
    near $\B$ but then becomes positive so that the light-sheet $\L$ of $\B$
    is incomplete. Points in the white region are the most interesting ones
    for testing the entropy bound. For $(\tau_i,\tau_f)$ in this region,
    light-sheets of surfaces $\B$ are complete.}
    \label{fig1}
  \end{center}
\end{figure}

As noted in Sec. \ref{s3.1}, the LQC solution is extremely
close to the classical FRW solution once the matter density $\rho$
falls to about $1\%$ of the Planck density $\rho_{\pl}$. Since the
covariant entropy bound is respected in the classical solution in
this regime, it suffices to examine a neighborhood of the bounce
in which $\rho > 10^{-2}\, \rho_{\pl}$. In the conventions used
above, the bounce occurs at $\tau=0$, and the matter density is
approximately $1.86\times 10^{-3}\rho_{pl}$ at $\tau= \pm
4t_{\pl}$. Therefore, it is sufficient to examine the region
$\tau_i,\:\tau_f\in \left[-4t_{\pl},4 t_{\pl} \right]$.

Because $\tau_i$ is necessarily less than $\tau_f$, the allowed
range of these parameters excludes the striped region in Fig.
\ref{fig1}. For $(\tau_i, \tau_f)$ in the allowed region, consider
the (future) null cone of any point $p$ at $\tau=\tau_i$ till it
intersects the surface $\tau=\tau_f$ in a round 2-sphere $\B$. If
the point $(\tau_i, \tau_f)$ lies in the black region, the expansion
of the light rays is positive at $\B$, whence no portion of the null
cone is a light-sheet of $\B$. (Thus, all $\B$ in the black region
are future trapped surfaces.) Therefore the black portion is
irrelevant for the conjecture. If a point $(\tau_i, \tau_f)$ lies in
the grey area, then the expansion is negative near $\B$ but
becomes positive as we move further back in time. So, in this case
the light-sheet is only a portion of the light cone. An explicit
calculation shows that, because of this truncation, the light-sheet
$\L$ is too short for the entropy flux through it to violate the bound.
Thus, to test the entropy bound, the nontrivial portion of the
$(\tau_i, \tau_f)$ plane is just the white region. The corresponding
surfaces $\B$ have complete past light-sheets.

\begin{figure}[tb]
  \begin{center}
  \subfigure[]
      {\includegraphics[width=3.75in]{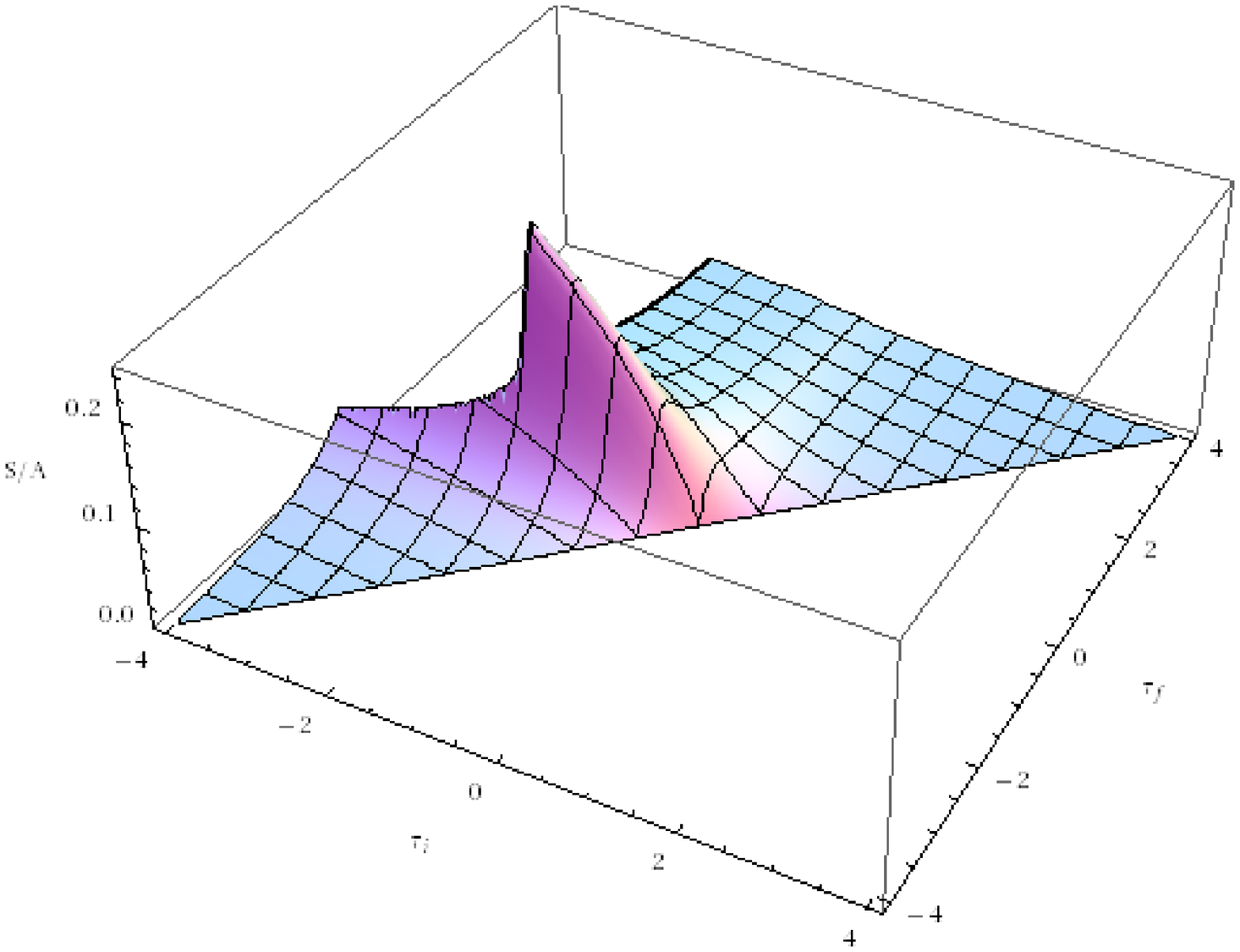}}
    \subfigure[]
      {\includegraphics[width=2in]{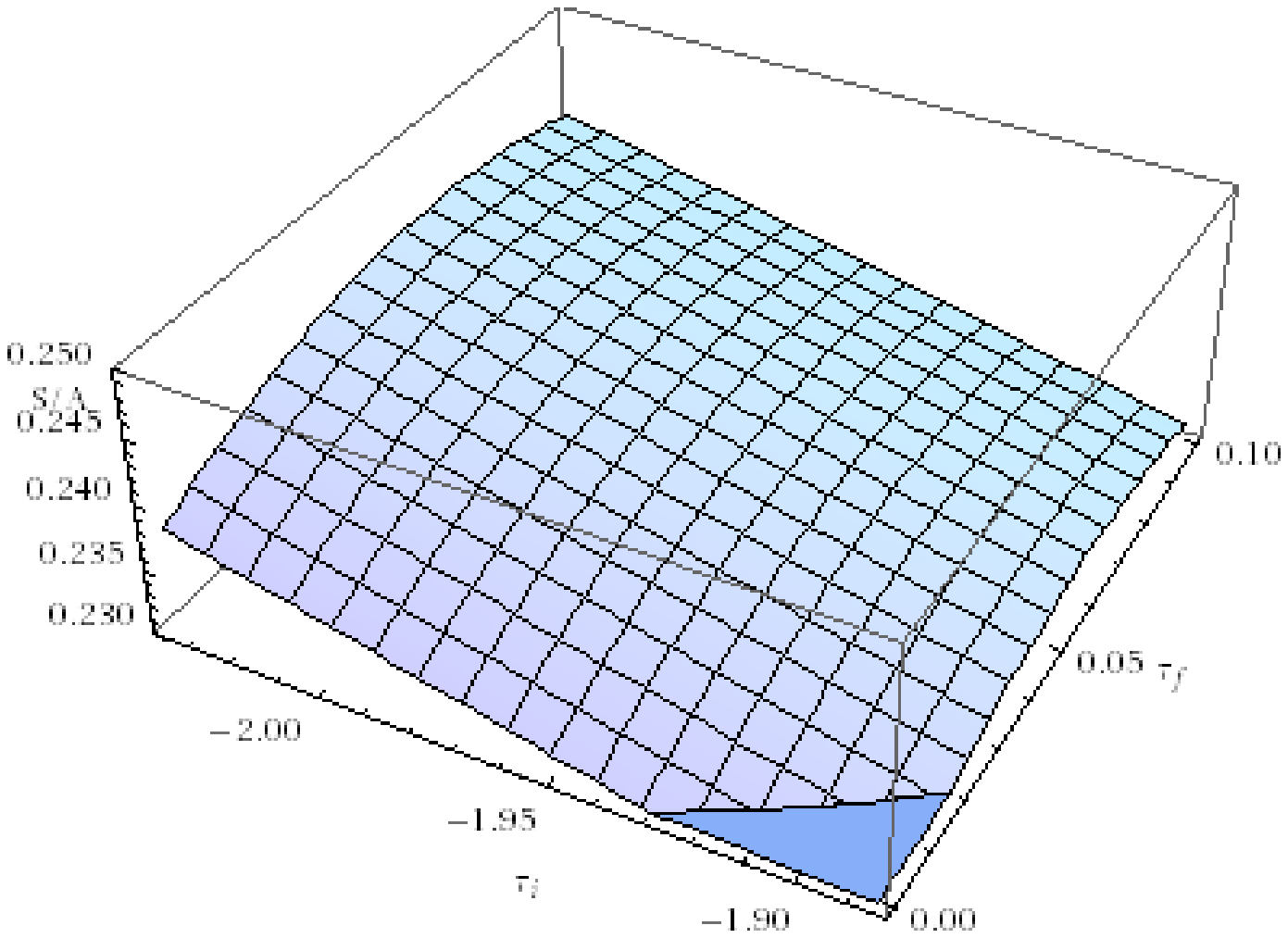}}
    \caption{$S/A$ is plotted in Planck units for points
     $(\tau_i,\tau_f)$ in the white region of Fig. 1.\\
     (a) Shows the full plot and (b) zooms in on the
    portion of the plot where $S/A$ is near its maximum.
    The maximum value is $S/A\, \approx \, 0.244/\lp^2$.}
  \label{fig2}
 \end{center}
\end{figure}

Figure \ref{fig2}a  shows the ratio $S/A$ in Planck units for
$(\tau_i,\tau_f)$ in this region, while Fig. \ref{fig2}b focuses on
the region where the ratio reaches its maximum.  The maximum
attained is
\be \f{S}{A}\,  \approx \, \f{0.244}{\lp^2}\, ,\ee
for $\tau_i\cong -2.035\,t_{\pl}$ and $\tau_f\cong0.0619\,
t_{\pl}$. Thus the ratio does approach the 1/4$\lp^2$ specified in
the covariant bound but does not quite reach it.

To conclude, let us return to the two questions raised earlier in
this subsection. In the above calculations we used the specific
values of $\gamma= 0.2375$ and $\a_o = 5.166$ for the parameters
$\gamma$ and $\a_o$ that appear in the expression of the quantum
corrected space-time metric. The value of $\gamma$ is well motivated
by black hole entropy considerations \cite{abk} but an independent
check is still lacking. The basis for the value of $\a_o$ is more
tentative, because we do not have a systematic procedure to arrive at
the Hamiltonian constraint of LQC from a specific, well-defined
proposal in LQG. However, since the quantum corrected geometry is
well defined for any value of these parameters, it follows that
$S/A$ would remain bounded even if these values were to shift. Thus,
the affirmative answer to the first question is robust. The second
question, on the other hand, refers to specific values of the
parameters. The fact that the bound is satisfied but almost
saturated near the bounce brings out an unforeseen coherence,
thereby providing independent circumstantial evidence that the
current values of $\gamma$ and $\alpha_o$ may be correct to a good
approximation.

\section{Discussion}
\label{s4}

In Sec. \ref{s2} we found that in the classical, radiation
filled FRW model, the covariant entropy bound is violated
near the big bang singularity. However, the violation occurs in the
region where space-time curvature and matter density are of Planck
scale. Therefore one would expect quantum gravity effects to be
dominant. The question is whether they can protect the bound. A
priori this appears to be a very difficult question to address
because quantum gravity effects would typically introduce large
fluctuations of geometry in the Planck regime, making the notion of
a light-sheet $\L$ of a spacelike surface $\B$
---and hence the very statement of the bound--- ill defined.

LQC provides a near ideal setting to investigate this issue
because of two features. First, the singularity is resolved:
quantum states can be evolved ``through'' the putative singularity.
The universe simply undergoes a quantum bounce and LQC equations
provide a deterministic evolution from a pre-big-bang branch to
our current post-big-bang branch. In FRW models, these results are
robust: they hold irrespective of whether there is a cosmological
constant \cite{bp}, whether we have a k=0 or k=1 universe
\cite{aps2,apsv}, and the singularity resolution is valid for all
states \cite{acs}. Second, the LQC wave function remains sharply
peaked on a smooth geometry even near the bounce point. Therefore,
there is a quantum corrected, ``effective'' metric ---obtained by
taking expectation values--- which reproduces the full quantum
dynamics to an excellent degree of approximation. Although its
coefficients involve $\hbar$, the metric is smooth and can be used
to introduce spacelike surfaces $\B$ and their light-sheets $\L$.
So, we can now ask: Is the covariant entropy bound respected in
this quantum corrected geometry? We cannot simply use one of
conditions \cite{fmw,bfm,ap} that ensures the satisfaction of the
bound because Einstein's equations do not hold on the quantum
corrected space-time. However, a direct calculation in Sec.
\ref{s3.2} showed that the answer is in the affirmative.

While this calculation provides an interesting convergence of
very different ideas related to quantum gravity, it has some
important limitations that stem from the current formulation of the
bound. First, the statement of the entropy bound requires a
well-defined notion of entropy current $s^a$. In practice this means
that the matter be described in hydrodynamical terms. Matter
described by ``fundamental'' classical and quantum fields does not
readily admit such a description. Furthermore, even in the
hydrodynamic context, entropy current is typically well-defined only
in equilibrium. These restrictions led us to use radiation fluid as
matter and assume that it is in instantaneous equilibrium throughout
the history of the universe. These approximations are suspect
especially in the Planck regime. The fact that the entropy bound
persists in spite of these seemingly crude approximations is
intriguing. Indeed even in the classical Friedmann universe, the
bound is respected after $\tau \approx 0.06 \tau_{\pl}$, a time at
which the matter density is about $8.3 \rho_{\pl}$. Why does the
bound continue to hold in such circumstances which are clearly
beyond the scope of approximations that are made? Such
``unreasonable'' successes have led to the suggestion that the bound
may have truly fundamental significance and should be a cornerstone
of any quantum gravity theory.

The most serious limitation of the current formulation of the bound
is that it requires a smooth classical geometry and, even on such a
geometry, one does not readily allow quantum matter fields. As
discussed in Sec. \ref{s1}, since quantum matter violates the
dominant (or null) energy condition, the available proofs break
down. Can one somehow incorporate quantum violations of energy
conditions? In the context of 2-dimensional, semiclassical
space-times which include the back reaction of the Hawking
radiation, there is an interesting proposal: modify the right-hand
side of the bound by addition of a term corresponding to
``entanglement entropy'' which could protect the bound in spite of
quantum violations \cite{st}. Perhaps the arguments available in the
classical theory \cite{fmw,bfm,ap} can be generalized to prove a
version of an appropriately modified bound also in four dimensions.
However the next step, dropping reference to a smooth classical
geometry, would be \emph{significantly} more difficult. What is to
become of the light-sheet $\L$ or indeed of a spacelike surface
$\B$ in situations \cite{aa-large,atv} in which, unlike quantum
cosmology, there are large fluctuations of geometry ?

Our overall viewpoint on the bound can be summarized as follows. The
bound is strongly motivated by the generalized second law (which
also does not have a well-defined, definitive formulation). Now,
already the standard second law of thermodynamics is a deep fact of
Nature but it has a ``fuzziness'' which is not shared by other deep
laws such as the conservation of energy-momentum and angular momentum.
In particular, the second law requires a coarse graining in an
essential way. It is not a statement about the evolution of micro-states;
in a fundamental theory their dynamics is always time reversible
(leaving aside, for simplicity, quantum measurements). Rather, it is
a statement about how the number of micro-states compatible with a
prespecified coarse graining changes in time. For instance if we
have a gas that is confined to the left half of a box and we open
the partition and wait till equilibrium is again reached, the
entropy increases. Any one of the final micro-states (occupying the
entire box) of these experiments will, under time-reversal, evolve
back to a micro-state that is confined to the left half.
Nonetheless, entropy increases because the \emph{total} number of
states compatible with the final coarse graining ---most of which
would not have resulted as end points of an \emph{actual} evolution
in our experiment--- is enormously larger than that corresponding to
the initial coarse graining. Thus, while an increase of entropy can be
calculated using statistical mechanics, it has little relevance to
the fundamental dynamics of micro-states. It is also not an input in
the construction of statistical mechanics. In the same vein, we
believe that the covariant entropy bound ---and its appropriate
generalizations that could encompass quantum field theory processes
even on ``quantum corrected'' but smooth space-times--- should
\emph{emerge} from a fundamental quantum gravity theory. These
bounds would be valuable but are not essential ingredients in the
\emph{construction} of such a theory. The distinguishing feature
about LQC, for example, is the underlying quantum geometry. The
covariant entropy bound was never an input or even a motivation.
Nonetheless, it emerged on the quantum corrected space-time as a
consequence of LQC.

\section*{Acknowledgements:} We would like to thank Martin Bojowald,
\'Eanna Flanagan, Ian Hinder, Jainendra Jain, Don Marolf, Tomasz
Pawlowski and Parampreet Singh for helpful discussions. This work
was supported in part by the NSF Grant No. PHY04-56913, The Natural Sciences
and Engineering Research Council of Canada and Le Fonds
qu\'eb\'ecois de la recherche sur la nature et les technologies, the
Alexander von Humboldt Foundation, and the Eberly research funds of
Penn State.

\end{document}